\magnification 1200 \hoffset-1truemm 

\def\nin{\noindent}
\def\d${$\displaystyle}
\def\ds{ \displaystyle}
\def\ut#1{$\underline{\smash{\hbox{#1}}}$}
\def\th{\vartheta} \def\rh{\varrho} \def\ph{\varphi}
\def\Ph{{\mit\Phi}}
\def\={\ = \ }
\def\b{\ }
\def\vg{v_{_{\ds{g}}}}
\def\r#1{r_{_#1}}
\def\parqua#1#2#3
 {\left[\matrix{
\kern-2pt{#1}&&\kern-20pt{#2}\cr&\kern-12pt{#3}&\cr}\kern-8pt
 \right]}
\def\parchr#1#2#3
 {\left\{\matrix{
\kern-2pt{#1}&&\kern-20pt{#2}\cr&\kern-12pt{#3}&\cr}\kern-8pt
 \right\}}
\def\Gammas#1#2#3{{\mit\Gamma}^{#1}_{{#2}{#3}}}
\def\arz{{a\over{r}_{_0}}}
\def\arzd{{\ds{a}\over\ds{r}_{_0}}}
\def\Dphialpha{\Delta\ph_{_{\ds\alpha}}}
\def\rho{\sqrt{r_0^2+a^2}}
\def\arrhoq{\left({ar_0\over{r_0^2+a^2}}\right)}

\line {} \vskip1truecm

\centerline{\bf On the gravitational field} \medskip \centerline{\bf A
suggestion about a possible experimental research} \bigskip

\centerline{ by B. Ferretti} \centerline{Dipartimento di Fisica
dell'Universit\`a di Bologna, Italy} \bigskip

{\bf Abstract}. {\it After a preliminary discussion of the relevance of the
field nature of gravitation interaction, both for the fundamental interaction of
particles and the topology of space time, a method is proposed to produce and
detect a dynamical gravitational field, allowing the determination of the order
of magnitude of its propagation velocity.} \bigskip

\item{\bf I} {\bf Introduction} \medskip

From the actual point of view of the physicist the structure of space and time
is merely depending on two fields: the electromagnetic and the gravitational
fields. On the other hand mathematically electromagnetism is entirely described
inside the Poincar\'e groups (enlarged Lorentz group), i.e. in the ambit of the
special relativity, whereas the gravitational field requires the consideration
of the larger set of transformation of general relativity. \medskip

For this and others more subtle reasons our knowledge of electromagnetism from
one side and of gravitation from the other presents itself in a quite different
way. \medskip

The very notion of field has been introduced by Faraday in the ambit of his
theory of electromagnetism, and formalized then by Maxwell. Electromagnetic
waves as such have then been produced and seen in laboratory by Hertz.
Eventually it has been seen that electromagnetic waves of Hertz's length are
produced in nature and that such waves fill the universe. \medskip

In the meantime a lot of very important things were happening in Physics: the
atomic theory of matter was put on new more solid basis, the electron and the
nuclei of the atoms were discovered and most important quantization was
discovered. \medskip

To the purposes of our discussion we have to notice that the only force which
had to be considered to give a completely (with the exclusion of few ``minor"
details) satisfactory theory of the structure of the atoms, molecules, solids
and so on was the electromagnetic force. \medskip

Even later, when nuclear physics and physics of elementary particles were
becoming of uttermost importance, new forces and new fields were hypothesized,
but always neglecting the gravitational forces. Gravitation was considered
instead only in the ambit of cosmology, astronomy and geophysics. \medskip

There was however the fundamental progress made by Einstein (see Ref. [1] and
[1']), who introduced the new idea that gravitational forces and inertial forces
were manifestations of the same field. This new idea allowed the generalization
of the special relativity. This generalization was on the other hand badly
needed to allow the consistency of relativity with many important facts of
physics. \medskip

The introduction of general relativity had no immediate effects on high energy
physics. \medskip

In this ambit the Heisenberg S matrix theory, with the very valid help of the
renormalization techniques, was of paramount importance. With this approach
questions about what really was happening at very small distances were
considered not well defined. Unfortunately, let us say it incidentally, even
what we should consider ``a small distance" was not well defined. \medskip

But even with this simplification, difficulties (for instance about not
renormalizability of certain fields) were persisting, and different S matrix
models had to be considered. \medskip

Between these models was Veneziano{'s} model (see Ref. [2]) in which, as far as
I know, for the first time an infinite number of ``elementary particles" was
introduced; furthermore the masses of these particles were going up to infinite.
On the other hand, as in that time usual, gravitational interaction was
completely neglected. I noticed almost immediately that the model couldn't be
consistent, because an elementary particle had to be considered as having the
simplest structure, and if its mass was $m$, its radius had to be $ r_{_C}
\simeq \hbar / mc $ (Compton's radius). Let $K$ be the universal gravitation
constant. The gravitational self energy of an elementary particle of mass $m$
should then be:

$$ E_{_K} \= {K m^2 \over r_{_C}} \= {K m^2 \over \hbar / mc } \= {Kc \over
\hbar} \ m^3 \ . $$

\nin Therefore the gravitational self energy $E_{_K}$ for an elementary particle
had to increase with the cube of the mass: that was meaning that gravitation
interaction couldn't any more be neglected in the theory of elementary
particles. \medskip

On the contrary, with a sufficiently high mass the gravitational self energy
becomes equal to the total self energy of the particle. That happens if

$$ mc^2 \= {Kc \over \hbar} \ m^3 \ , \hskip1truecm {\rm i.e. \ if}
\hskip1truecm m \= \sqrt{\hbar c \over K} \ \simeq \ 2\times10^{-5} \ {\rm{g}} \
. $$

The ``elementary particle" therefore becomes a very small black hole. It should
be noticed that the corresponding Compton's radius is

$$ r_{_C} \= {\hbar \over mc} \= {10^{-27} \over 2 \cdot 10^{-5} \ \times \ 3
\cdot 10^{-10}} \ \simeq \ 2\times10^{-33} \ {\rm{cm}} \ . $$

This distance

$$ r_{_P} \= \sqrt{\hbar K \over c^3} $$

\nin had been introduced in physics, about fifty years before I ``rediscovered"
it, by Planck who wanted to form three ``natural" units of length, time and mass
by mean of universal constants only, whereas I was thinking about the elementary
particles. \medskip

\d$ r_{_P} $ was merely the ``natural unity of length" of Planck. I confess that
just I ignored the thing. In that time I was working in the theoretical division
of CERN. Of course I started to talk with my colleagues theoreticians of my
reflexion about the model of Veneziano. Now to my knowledge nobody had mentioned
the idea that gravity might be important in the theory of elementary particles
up to that time; after one year it was common place. \medskip

On my side my attention was particularly attracted by the idea that gravitation
might be very important to determine the structure of space-time and not only at
great distances as suggested by the general relativity theory, but also at very
small distances determining there the space-time topological structure. \medskip

I started then to investigate theoretically which might be the influence of
gravitation on high precision measurements at very small distances (see Ref.
[3], [4], [5]). During this investigation I convinced myself that it was
probably essential to consider, as in general relativity, gravitation as a field
of forces. On the other hand I was more and more impressed by the fact that the
direct experimental evidence of this assumptions is up to now missing. \medskip

During the second half of the last century a search of the gravitational waves
coming from space up to now has not given any sure positive results (see Ref.
[6]). \medskip

Recently evidence of a finite velocity propagation of gravitation was obtained
by astronomical measures (see Ref. [7], [8]). \medskip

On the other hand, due to the smallness of the gravitational constant, the
enterprise of detecting the propagation of a dynamical (i.e. varying in time)
gravitational field in the laboratory, like made by Hertz on the electromagnetic
waves, appears to be rather difficult. \medskip

One might even wonder whether it is worthwhile. What could we learn from it?
Should for instance we be able to measure their velocity of propagation, might
we expect something different from the velocity of light? For restricted
relativity reason, that would be unthinkable. But for this very reason, should
we be able to measure the finite velocity of propagation of the gravitational
field, it would be imperative to do it. \medskip

I decided therefore to investigate the possibility of what I have called an
Hertzian enterprise about gravitation. This is the object of the present paper.
\bigskip

\item{\bf II \phantom{A}} {\bf A suggestion about the possibility of producing
and revealing gravitational waves allowing the determination of the order of
magnitude of their propagation velocity} \bigskip

\item{\bf II A} {\bf The general scheme of the suggested experiment} \medskip

The experimental apparatus will include a certain number of ``radiators" of
gravitation and of ``detectors". In general the radiators will be rotating
bodies and the frequency of rotation will be the same for all \footnote{$(^1)$}
{\sl My attention was called on two interesting papers (see Ref. [9], [10]) in
which use was made of rotating radiators of gravity. The aim of these papers was
a search of a force due to a hypothetical Yukawian potential superposed to the
usual Newtonian potential. The frequency of the rotators, the masses and the
distances were far different from those which I have considered in the present
work, and therefore I was unable to use their interesting results to modify the
present paper.}. As a consequence, the gravitational potential in the space
surrounding the radiators will be, as a function of time, analyzable in Fourier
series. \medskip

If on the other hand the detector is finely tuned on the frequency of the
considered Fourier component, it becomes possible to get a ``zero measurement"
of the velocity of propagation. \medskip

This is the fundamental idea underlaying our suggestion. \bigskip

\item{\bf II B} {\bf The expression of the gravitational field gravity following
general relativity in the limit of small velocity of the gravitating bodies}
\medskip

Let $$ ds^2 \= \sum_{\mu,\nu} g_{\mu\nu} dx_{\mu} dx_{\nu} \leqno(B1) $$

\nin be the metric of space-time in the reference [1] and [1']. Let be (see Eq.
21 of Ref. [1']):

$$ \parqua{\mu}{\nu}{\sigma} \= {1\over 2} \ \left[
{\partial{g}_{\mu,\sigma}\over\partial{x}_\nu}
+{\partial{g}_{\nu,\sigma}\over\partial{x}_\mu}
-{\partial{g}_{\mu,\nu}\over\partial{x}_\sigma} \right] \ , \leqno(B2) $$

\nin and (see Eq. 23 of Ref. [1']):

$$ \parchr{\mu}{\nu}{\gamma} \= \sum_\alpha g^{\gamma\alpha}
\parqua{\mu}{\nu}{\alpha} \leqno(B2') $$

\nin (Christoffel symbols). \medskip

The equation of motion of the material point in $K$ are given by (see Eq. 46 of
Ref. [1']):

$$ {\partial^2x_\gamma\over ds^2} \= \Gammas{\gamma}{\mu}{\nu}
{\partial{x}_\mu\over ds} {\partial{x}_\nu\over ds} \ , \leqno(B3) $$

\nin where \d$ \Gammas{\gamma}{\mu}{\nu} \= - \parchr{\mu}{\nu}{\gamma} \ . $
\medskip

The preceding equations are the general relativity equations of motion. In weak
field approximation, as certainly we have to be, the metric tensor is given by:

$$ g_{\mu\nu}^{W} \= \left(\matrix{
-1+\delta{g}_{_{11}}&
   \delta{g}_{_{12}}&
   \delta{g}_{_{13}}&
   \delta{g}_{_{14}}\cr
   \delta{g}_{_{21}}&
-1+\delta{g}_{_{22}}&
   \delta{g}_{_{23}}&
   \delta{g}_{_{24}}\cr
   \delta{g}_{_{31}}&
   \delta{g}_{_{32}}&
-1+\delta{g}_{_{33}}&
   \delta{g}_{_{34}}\cr
   \delta{g}_{_{41}}&
   \delta{g}_{_{42}}&
   \delta{g}_{_{43}}&
+1+\delta{g}_{_{44}}\cr}
\right) \ , \leqno(B4) $$

\nin where \d$\delta{g}_{\mu\nu}$ denote small quantities. \medskip

The special relativity metric tensor is given by:

$$ g_{\mu\nu}^{S} \= \left(\matrix{
-1&0&0&0\cr
0&-1&0&0\cr
0&0&-1&0\cr
0&0&0&+1\cr} \right) \leqno(B4') $$

Consequently the difference of \d$g_{\mu\nu}$ from \d$g_{\mu\nu}^{S}$ (see B4
and B4') will be considered small (indeed extremely small) of the first order
(and their square generally neglected), and in a similar way
\d${\partial{x}_k\over dx_4}$ will be considered small of the first order.
\medskip

With this approximation, the equation of motion (B3) becomes, writing
\d$dt=ds=dx_4$ (see Eq. 67 of Ref. [1']):

$$ {\partial^2x_k\over dt^2} \= - {1\over 2}
{\partial{g}_{44}\over\partial{x}_k} \hskip2truecm k=1,2,3 \ \ ; \leqno(B3') $$

\nin which implies, in order to get the Newton limit:

$$ {1\over 2} {g}_{44} \= \Ph \ , \leqno(B5) $$

\nin where $\Ph$ is the Newton gravitational potential. \medskip

This is our first fundamental relation. To get now the relation between $\Ph$
and the density of matter $\rh$, we have to consider the general relativity
field equations. \medskip

The field equations are (see Eq. 57, 58, 58a, 58b of Ref. [1']):

$$ \sum_\alpha {\partial\Gammas{\alpha}{\mu}{\nu}\over{\partial{x}}_\alpha} +
\sum_{\alpha,\beta} \Gammas{\alpha}{\mu}{\beta} \Gammas{\beta}{\nu}{\alpha} \=
-\kappa \left(T_{\mu\nu}-{1\over 2}{g}_{\mu\nu}T\right) \ , \leqno(B6) $$

\nin where \d$ \kappa \= {8\pi{K}\over{c^2}}, $ \ $T$ is the trace of
\d$T_{\mu\nu}$, \ \d$ \sqrt{-g} \= 1, $ \ and \d$ g \= \left|g_{\mu\nu}\right| \
. $ \medskip

In our approximations \d$\Gammas{\alpha}{\mu}{\beta}$ are small of the first
order and therefore \d$\sum_{\alpha,\beta} \Gammas{\alpha}{\mu}{\beta}
\Gammas{\beta}{\nu}{\alpha}$ at the first member of equations (B6) can be
neglected. \medskip

On the other hand, always with our approximation (weak field approximation), we
have:

$$ \sum_\alpha {\partial\Gammas{\alpha}{\mu}{\nu}\over\partial{x}_\alpha} \=
{\partial\over\partial{x}_1} \parqua{\mu}{\nu}{1} + {\partial\over\partial{x}_2}
\parqua{\mu}{\nu}{2} + {\partial\over\partial{x}_3} \parqua{\mu}{\nu}{3} -
{\partial\over\partial{x}_4} \parqua{\mu}{\nu}{4} \ . \leqno(B7) $$

Now, if we put in the preceding formula $\mu=\nu=4$, we get (see B2):

$$ \sum_\alpha {\partial\Gammas{\alpha}{4}{4}\over{\partial{x}}_\alpha} \=
-{1\over 2} \left[ \nabla^2g_{44} + {\partial^2\over\partial{x}^2_4}g_{44}
\right] \ . \leqno(B8) $$

It is very important to notice that in the original paper by Einstein about the
general relativity (see Eq. 68 of Ref. [1']) the term
\d$-{1\over2}{\partial^2\over\partial{x}^2_4}g_{44}$ in the passage to the
Newtonian limit was omitted. That was perfectly consistent with the
approximations performed to the purpose. In our case however, as we shall see,
the term in question is essential. \medskip

In the passage from equation (B7) to equation (B8) we neglected

$$ \sum_{\ell=1}^3 {\partial\over\partial{x}_{_\ell}}
{\partial\delta{g}_{\ell4}\over\partial{x}_4} \ , $$

\nin as the \d$\delta{g}_{\ell4}$ are presumably very small. Indeed all
components of the gravitational field different from \d$g_{44}$ are always
neglected in all astronomical calculations. Shouldn't be correct, just an
experiment as that we suggest would put it in evidence. \medskip

On the other hand, considering again the equation (B6), \d$T_{\mu\nu}$ is the
tensor of energy and momentum of the gravitation field and of the matter.
Eventually we get for the gravitational potential the equation:

$$ {1\over 2}\nabla^2{g}_{44} - {1\over 2} {1\over{c}^2}
{\partial^2\over\partial{t}^2_4}g_{44} \= {1\over 2} \kappa\rh \ , \leqno(B9) $$

\nin where $\rh$ is the density of the gravitating matter and \d$ \ {1\over 2}
\kappa = \ {4\pi K \over c^2} \ , $ where $K$ is the Newtonian gravitational
constant and has the value \d$ 6.673 (10) \times 10^{-8} $ in C.G.S. units
\d$\left( {\rm{cm}}^3 \ {\rm{g}}^{-1} \ {\rm{s}}^{-2} \right)$. \medskip

Now, due to the term \d$ {1\over 2} {1\over{c}^2}
{\partial^2\over\partial{t}^2_4}g_{44} \ , $ the solution of (B9) cannot anymore
be the Newton gravitational potential, but for instance:

$$ \Ph(t-r/\vg) \= -K \ \int {\rh(x',y',z',t-r/\vg)\over{r}} \ dx' dy' dz' \ ,
$$

\nin \ut{supposing} \ut{that} \ut{the} \ut{velocity} \ut{of} \ut{propagation}
\ut{is} $\vg$. \medskip

\d$\Ph(t-r/\vg)$ differs from the former usual Newton potential by a correction
of the time in the density \d$\rh(x',y',z',t')$ of gravitating matter for the
retardation due to the velocity of propagation. For consistence with Eq. B9,
this velocity \d$\vg$ is expected to be just the velocity of light. \medskip

The chief object of the suggested research should be to put in evidence this
correction, and it is just a correction of this kind which characterizes fields
against action at distance. \bigskip

\item{\bf II C} {\bf The simple radiator} \medskip

Schematically our radiator consists of two equal bodies
\d${\cal{R}}{_1},{\cal{R}}{_2}$ of equal masses \d$M_{_R}$ made of the same
substance, for instance iron, having the same geometrical form (sphere), with
the centers of gravity laying in \d${\cal{G}}_1$ and \d${\cal{G}}_2$ on the same
horizontal plane at a distance $2a$ and rotating around a vertical axis passing
by the middle point $O$ of \d${\cal{G}}_1,{\cal{G}}_2$. We suppose that $a$ is
much greater than the radius of the spheres \d${\cal{R}}{_1},{\cal{R}}{_2}$ \ .
\medskip

The frequency of rotation will be $\nu$ hertz (turns by second), and
consequently the centrifugal force acting on each body and due to the rotation
will be:

$$ {\cal{F}}_{_C} \= (2\pi\nu)^2 \int\kern-6pt\int\kern-6pt\int \sigma
{\xi^2+\eta^2 \over \sqrt{\xi^2+\eta^2}} \ d\xi d\eta d\zeta \ \simeq \
(2\pi\nu)^2 M_{_R} a \ , $$

\nin where $\sigma$ is the density of the rotating body, $\xi,\eta,\zeta$ the
Cartesian coordinates ($\xi,\eta$ the horizontal, $\zeta$ the vertical) of the
particles of the body, and $a$ must be much greater than the linear dimension of
the body. \medskip

For instance, if $a=100$ cm, \d$M_{_R}=1000$ kg \d$=10^6$ gr, \d$\nu=1$ hertz:

$$ {\cal{F}}_{_C} \ \simeq \ \left( (2\pi)^2 \times 10^6 \times 100 \right) \
{\rm{dyn}} \ \simeq \ 4 \times 10^9 \ {\rm{dyn}} \= 4 \times 10^4 \ {\rm{N}} \ .
$$

The centrifugal forces are equal and opposite and the equilibrium shall be kept
by a tension bundle of steel wires, or, to get a symmetrical arrangement, a ring
of steel and carbon fibers. \medskip

We can use cylindrical coordinates $\rh,\ph,\zeta$ of center $O$, (middle point
of \d${\cal{G}}_1,{\cal{G}}_2$), being $\xi=\rh\cos\ph$, $\eta=\rh\sin\ph$ and
$\ph=2\pi\nu{t}+\psi/2$. The dependence on time is totally contained in $\ph$.
The other parameters are time independent. \medskip

Now, if $x,y,z$ are the Cartesian coordinates of a given point $P$ (for instance
of the detector), the distance between $P$ and $Q$ (a point of the rotating
body) is given by:

$$ r \= \sqrt{ (x-\xi)^2 + (y-\eta)^2 + (z-\zeta)^2 } \ . $$

We assume that the points $P$ all belong to a small neighbourhood of a point
\d$P_0$ of coordinates

$$ x_{_0} \= \r0 \ , \hskip1truecm y_{_0} \= 0 \ , \hskip1truecm z_{_0} \= 0 \ ;
\leqno(C1) $$

\nin and therefore

$$ x \= \r0 + \Delta x \ , \hskip1truecm y_{_0} \= 0 + \Delta{y} \ ,
\hskip1truecm z_{_0} \= 0 + \Delta{z} \ ; $$

\nin where

$$ {\Delta{x}^2+y^2+z^2\over{\r0}^2} \ \ll \ 1 \ . \leqno(C1') $$

Similarly \d$\xi=a\cos\ph+\Delta\xi$, \d$\eta=a\sin\ph+\Delta\eta,$
\d$\zeta=0+\Delta\zeta$. \medskip

It follows that we can write

$$ r^2 \= \left(\r0 - a \cos\ph + \Delta{x} - \Delta\xi\right)^2 + \left(-a
\sin\ph + \Delta{y} - \Delta{\eta} \right)^2 + \left(\zeta-\Delta{z}\right)^2 \
= $$ $$ \phantom{r^2} \= \r0^2 \left\{ \left(1 - \arz\cos\ph +
{\Delta{x}-\Delta\xi\over \r0} \right)^2 + \left( - \arz\sin\ph +
{\Delta{y}-\Delta\eta\over \r0} \right)^2 + \left( {\zeta-\Delta{z} \over \r0}
\right)^2 \right\} \ . $$

Supposing that the inequality (C1') is very strong, we can write:

$$ r^2 \= \r0^2 \left\{ \left(1-\arz\cos\ph\right)^2 +
\left(\arz\sin\ph\right)^2 + {\rm small \ terms} \right\} \leqno(C2) $$

\nin and neglecting the small terms, eventually we get:

$$ r^2 \= \r0^2 \left\{ \left(1-2\arz\cos\ph\right) +
\left({a\over{r}_{_0}}\right)^2 \right\} \ . \leqno(C3) $$

We notice that, if the rotating bodies \d${\cal{R}}{_1}$ and \d${\cal{R}}{_2}$
are regular spheres and the mass of the bundles are negligible, corrections due
to \d$\Delta\xi,$ \d$\Delta\eta,$ \d$\Delta\zeta$ are exactly zero. \medskip

Now our radiator consists schematically of two bodies
\d${\cal{R}}{_1},{\cal{R}}{_2}$, and if for \d${\cal{R}}{_1}$ we have:

$$\cases{
\ph \hskip1truemm \= \ph_1 \= 2\pi\nu{t}+\psi/2 \ , \cr\cr
\r1^2\=\r0^2\left\{ 1-2\arzd\cos\left(2\pi\nu{t}+\psi/2\right) +
\left({\ds{a}\over\ds{r_{_0}}}\right)^2 \right\} \ ; \cr } $$

\nin correspondingly for \d${\cal{R}}{_2}$ we have:

$$\cases{
\ph_2 \hskip1truemm \= \ph_1+\pi \ , \cr\cr
\r2^2\=\r0^2\left\{1+2\arzd\cos\left(2\pi\nu{t}+\psi/2\right) +
\left({\ds{a}\over\ds{r_{_0}}}\right)^2 \right\} \ . \cr } $$

We can also write:

$$\cases{
\r1^2\=\r0^2-2a{\r0}\cos\ph+a^2 \ , \cr\cr
\r2^2\=\r0^2+2a{\r0}\cos\ph+a^2 \ , \cr\cr
\ph\hskip1truemm\=2\pi\nu{t}+\psi/2 \ . \cr } \leqno(C3') $$

The gravitational potential $\Ph(x,y,z,t)$ created by the radiator in $P$ (with
our approximation and neglecting for the moment the retardation) is given by

$$ \Ph \= -KM_{_R} \left( {1\over{\r1}} + {1\over{\r2}} \right) \ . \leqno(C4)
$$

In order to calculate $\Ph$ as given by (C4) we could use the expansion in
multipoles, as commonly in potential theory, or, in an equivalent way, the
expansion in powers of $\xi$ of the function $1/\sqrt{1+\xi}$ \ :

$$ \left(1+\xi\right)^{-1/2} \= 1 - {1\over 2}\xi + {3\over2^2} {\xi^2\over2!} +
\ \dots \ + (-1)^n {1\over2^n} \left(\prod_{l=1}^{n-1}(2\ell+1)\right)
{\xi^n\over{n}!} + \ \dots $$

We can write:

$$ r_{1,2} \= \ \sqrt{\r0^2\mp2ar_0\cos\ph+a^2} \=
\ \rho \ \sqrt{1\mp2\arrhoq\cos\ph} \ , $$

\nin so, in our case:

$$ \xi_{1,2} \= \mp \ 2\arrhoq\cos\ph \ . $$

\nin It follows therefore:

$$ {1\over{\r1}} + {1\over{\r2}} \= {2\over{\rho}} \left\{ 1 + \sum_{m=1}^\infty
{1\over(2m)!} \left(\prod_{\ell=1}^{2m-1}(2\ell+1)\right) \arrhoq^{2m}
\cos^{2m}\ph \right\} \ . \leqno(C5) $$

Now, as we have suggested in IIA, we want to use the Fourier expansion of the
gravitational potential. But in (C5) the time is contained only in the factors
\d$\cos^{2m}\ph$ and therefore we have merely to take the Fourier expansion of
$\cos^{2m}\ph$:

$$ \cases{
m\=1: & \d$
\cos^2\ph \= {\ds{1}\over\ds{2}} + {\ds{1}\over\ds{2}}\cos2\ph \ , $ \cr\cr
m\=2: & \d$
\cos^4\ph \= {\ds{3}\over\ds{8}} + {\ds{1}\over\ds{2}}\cos2\ph +
{\ds{1}\over\ds{8}}\cos4\ph
\ , $ \cr\cr
m\=3: & \d$
\cos^6\ph \= {\ds{5}\over\ds{16}} + {\ds{15}\over\ds{32}}\cos2\ph +
{\ds{3}\over\ds{16}}\cos4\ph + {\ds{1}\over\ds{32}}\cos6\ph \ , $ \cr\cr
m\=4: & \d$
\cos^8\ph \= {\ds{35}\over\ds{128}} + {\ds{7}\over\ds{16}}\cos2\ph +
{\ds{7}\over\ds{32}}\cos4\ph + {\ds{1}\over\ds{16}}\cos6\ph +
{\ds{1}\over\ds{128}} \cos8\ph \ , $ \cr\cr
\dots \phantom{=} \dots & \d$ \dots \dots \dots \dots \dots \ . $ \cr\cr
} \leqno (C6) $$

and put them in (C5) to obtain through (C4) the desired expansion. \medskip

{\bf Remarks} \medskip

\item{\phantom{I}I)} We neglect all retardation effects \ut{inside} the
\ut{single} \ut{radiator}, in accordance with the assumption that the radiator
is ``small". The ``time" of the radiator is the ``time" of the ``center" of the
radiator. Therefore:

$$\cases{
\ph \= \ph_1 \= 2\pi\nu{t}+\psi/2 \ , \cr
\phantom{\ph \ \ } {\rm{or}} \cr
\ph \= \ph_2 \= \ph_1 + \pi \ ; \cr } $$

\nin where \d$\psi/2$ is the initial phase. \medskip

We observe that in (C5) are contained only even powers of $\cos\ph$. \medskip

As a consequence, the \ut{fundamental} frequency irradiated is not the rotation
frequency, but the double of such a frequency. This fact on the other hand is
obvious, remarking that the radiator regains its initial disposition after half
a tour. \medskip

A similar property is valid for a rotating regular polygon having identical
bodies at its vertexes and it is valid independently from retardation: the
fundamental frequency emitted is $n$ times the rotating frequency ($n$ number of
sides of the polygon). \medskip

\item{II)} If \d$a\ll\r0$, as we suppose, the series of (C5) converges quite
rapidly while $m$ is increasing: fully analytical calculations have been carried
out \footnote{$(^2)$} {\sl By S. Turrini, Dept. of Physics of the University of
Bologna, using program FORM by J.Vermaseren, see Ref. [11].} up to $m=15$, i.e.
up to \medskip

$$\arrhoq^{2m}\cos^{2m}\ph \= \arrhoq^{30}\cos^{30}\ph$$.

We write therefore:

$$ \Ph \= - {2KM_{_R}\over{\rho}} \left\{ 1 + {3\over2!}\arrhoq^2\cos^2\ph +
{105\over4!}\arrhoq^4\cos^4\ph + \right. $$ $$ \hskip2truecm \left. +
{10\b395\over6!}\arrhoq^6\cos^6\ph + {2\b027\b025\over8!}\arrhoq^8\cos^8\ph + \
\dots \right\} \ , \leqno(C7) $$

and then, using the Fourier expansion (C6) of $\cos^{2m}\ph$: \medskip

\nin \d$ (C8) \hskip1truecm \Ph \= - {2KM_{_R}\over{\rho}} \ \ \Bigg\{ \left[ +
1 + {1\over2}{3\over2!}\arrhoq^2 + {3\over8}{105\over4!}\arrhoq^4 + \right. $
\par \d$ \hskip42truemm \left. + {5\over16}{10\b395\over6!}\arrhoq^6 +
{35\over128}{2\b027\b025\over8!}\arrhoq^8 + \dots\ \right] $ \medskip \d$
\hskip28truemm + \cos2\ph \ \ \times \left[ + {1\over2}{3\over2!}\arrhoq^2 +
{1\over2}{105\over4!}\arrhoq^4 + \right. $ \medskip \d$ \hskip43truemm \left. +
{15\over32}{10\b395\over6!}\arrhoq^6 + {7\over16}{2\b027\b025\over8!}\arrhoq^8 +
\dots\ \right] $ \medskip \d$ \hskip28truemm + \cos4\ph \ \ \times \left[ +
{1\over8}{105\over4!}\arrhoq^4 + {3\over16}{10\b395\over6!}\arrhoq^6 + \right. $
\medskip \d$ \hskip53truemm \left. + {7\over32}{2\b027\b025\over8!}\arrhoq^8 +
\dots\ \right] $ \d$ \hskip5truemm + \ \dots \hskip5truemm \ \Bigg\} $ \medskip

As just told, terms up to \d$\arrhoq^{30}$ have been analytically taken in
account. \medskip

We want now to study the behaviour of our detector in any potential which can be
described in a Fourier series. Suppose then that the gravitational field is
given by:

$$ \Ph(x,y,z,t) \= \sum_{m=1}^\infty \Ph_m(x,y,z) \sin(2m\omega{t}+\sigma_m) \ ,
$$

\nin where the \d$\sigma_m$ are the initial phases. \medskip

The equation of motion of the detector will be (see sect. II D):

$$ M_{_D} \ddot{x} + gx \= \ - \ M_{_D} \sum_{m=1}^\infty
{\partial\Ph_m\over\partial{x}} \sin(2m\omega{t}+\sigma_m) \ . \leqno(C9) $$

If \d$x_{_{\ds\ell}}(t)$ is a particular solution of the equation

$$ M_{_D} \ddot{x} + gx \= \ - \ M_{_D} {\partial\Ph_\ell\over\partial{x}}
\sin(2m\omega{t}+\sigma_\ell) \ , $$

\nin given the linearity of the equation (C9)

$$ x(t)\=\sum_{\ell} c_{_{\ds\ell}} x_{_{\ds\ell}}(t) \ , $$

\nin where the \d$c_{_{\ds\ell}}$ are constant coefficients, is a particular
solution of (C9). \medskip

Now, for reasons which we shall see in the following, \ut{if}
\d$\omega\=\sqrt{g/M_{_D}}$ only the solution

$$ x_{_{\ds\ell}} \= x_{_{\ds1}} \hskip2truecm (\ell=1) $$

\nin will be interesting. \bigskip

\item{\bf II D} {\bf The detector coupling with the radiator} \medskip

The detector consists of a mass \d$M_{_D}$ coupled with a spring, whose recall
force to a very great mass at the equilibrium point $x=0$ is $-gx$. Let us write
the gravitational force on the detector as:

$$ {\cal{F}} \sin\omega{t} \= M_{_D} {\partial\Ph\over\partial{x}} \sin\omega{t}
\ . $$

We suppose now:

$$ \omega \= \sqrt{g \over M_{_D} } \ . \leqno(D1) $$

The equation of the motion of the detector without any field is:

$$ M_{_D}\ddot{x} + gx \= 0 \ . \leqno(D2) $$

The interesting general solution of the equation (C9) is therefore written as:

$$ x(t) \= \xi(t)\sin\omega{t} + \eta(t)\cos\omega{t} \ , \leqno(D3) $$

\nin and making the position:

$$ \dot\xi\sin\omega{t} + \dot\eta\cos\omega{t} \= 0 \ , \leqno(D4) $$

\nin we get:

$$ M_{_D} \ddot{x} + gx \= - M_{_D} \omega^2 x + M_{_D} \omega
\left(\dot\xi\cos\omega{t}-\dot\eta\sin\omega{t}\right) + gx \= M_{_D} \omega
\left(\dot\xi\cos\omega{t}-\dot\eta\sin\omega{t}\right) $$

\nin because of (D1). \medskip

Therefore we get the equations:

$$\cases{
\dot\xi\sin\omega{t} + \dot\eta\cos\omega{t} \= 0 \ , \cr \cr
\dot\xi\cos\omega{t} - \dot\eta\sin\omega{t} \= {\ds1\over\ds\omega}
{\ds\partial\Ph\over\ds\partial{x}} \sin\omega{t} \ . \cr } $$

It follows:

$$\cases{
\dot\xi \= + {\ds1\over\ds\omega} {\ds\partial\Ph\over\ds\partial{x}}
\sin\omega{t}\cos\omega{t} \ , \cr \cr
\dot\eta \= - {\ds1\over\ds\omega} {\ds\partial\Ph\over\ds\partial{x}}
\sin^2\omega{t} \ ; \cr } $$

\nin and therefore ($\tau\equiv\omega{t}$):

$$\cases{
\xi \= \xi_0 + {\ds1\over\ds\omega^2} {\ds\partial\Ph\over\ds\partial{x}}
\int_0^{\ds\omega{t}} \sin\tau\cos\tau d\tau \= \xi_0 + {\ds1\over\ds2\omega^2}
{\ds\partial\Ph\over\ds\partial{x}} \sin^2\omega{t} \ , \cr \cr
\eta \= \eta_0 - {\ds1\over\ds\omega^2} {\ds\partial\Ph\over\ds\partial{x}}
\left[ {\ds1\over\ds2}\omega{t} - {\ds1\over\ds4}\sin(2\omega{t}) \right] \ .
\cr } \leqno(D5) $$

Recalling (D3), \medskip

\d$ x(t) \= \xi\sin\omega{t} + \eta\cos\omega{t} \= $ \medskip \d$
\phantom{x(t)} \= \left\{ \xi_0 + {1\over\omega^2} {\partial\Ph\over\partial{x}}
\left(\sin\omega{t}\right)^2\right\} \sin\omega{t} \ + \ \left\{ \eta_0 -
{1\over\omega^2} {\partial\Ph\over\partial{x}} \left[ {1\over2}\omega{t} -
{1\over4}\sin(2\omega{t}) \right] \right\} \cos\omega{t} \ . $ \medskip

The only relevant term in \d$x(t)$ is: \medskip

$$ - {1\over\omega^2} {\partial\Ph\over\partial{x}} {1\over2}\omega{t}
\cos\omega{t} \= - {1\over2\omega} {\partial\Ph\over\partial{x}} t \cos\omega{t}
\ , $$ \medskip

\nin the other terms are oscillating terms and not increasing with time in
absolute value, so:

$$ x_{_1}(t) \= x(t) \= - {1\over2\omega} {\partial\Ph\over\partial{x}} t
\cos\omega{t} \ + \ \dots \ . \leqno(D6) $$ \medskip

The same holds for the velocity: \medskip

\d$ \dot{x}(t) \= \dot\xi\sin\omega{t} + \dot\eta\cos\omega{t}
+ \omega \left[ \xi\cos\omega{t} - \eta\sin\omega{t} \right] \=
+ \omega \left[ \xi\cos\omega{t} - \eta\sin\omega{t} \right] \ , $ \medskip

\nin due to (D3) and (D4), and then, due to (D5): \medskip

\d$ \dot{x}(t) \=
+ \omega \left\{ \xi_0 + {\ds1\over\ds2\omega^2}
{\ds\partial\Ph\over\ds\partial{x}} \sin^2\omega{t} \right\} \cos\omega{t} \ + $
\medskip \d$ \phantom{\dot{x}(t) \=}
- \omega \left\{ \eta_0 - {\ds1\over\ds\omega^2}
{\ds\partial\Ph\over\ds\partial{x}} \left[ {\ds1\over\ds2}\omega{t} -
{\ds1\over\ds4}\sin(2\omega{t}) \right] \right\} \sin\omega{t} \ . $ \medskip

As before, the only relevant term is:

$$ + \omega{1\over\omega^2} {\partial\Ph\over\partial{x}} {1\over2}\omega{t}
\sin\omega{t} \= + {1\over2} {\partial\Ph\over\partial{x}} t \sin\omega{t} \ ,
$$ \medskip

\nin as the other terms are oscillating terms and not increasing with time in
absolute value, so:

$$ \dot{x}(t) \= + {1\over2} {\partial\Ph\over\partial{x}} t \sin\omega{t} \ + \
\dots \ . \leqno(D7) $$

All that is valid until friction is neglected. \medskip

The kinetic energy of the detector,

$$ T \= {1\over 2} M_{_D} \dot{x}^2(t) \= {1\over 2} M_{_D} \ {1\over4} t^2 \
\left({\partial\Ph\over\partial{x}}\right)^2 \ \sin^2\omega{t} \ + \ \dots \ ,
\leqno(D8) $$

\nin might become important enough without necessity of going to extremely low
temperatures, as I think quite probable that it exists the possibility to reach
damping periods of the order or greater than \d$10^4$ sec. \medskip

Now it is certainly clear that the only interesting case is that of
\d$x_{_1}(t)$ (see D6), because only this case admits the ``resonant" solution
allowing sensibility impossible in other cases. \bigskip

\item{\bf II E} {\bf The measurement setup} \medskip

From equations (C8) it follows easily that the gravitational potential generated
by a simple radiator can be written: \medskip

\d$ (E1) \hskip5truemm \Ph \= - 2K{M\over{\rho}} \times \left\{ F_0(a,\r0) +
\right. $ \par \hskip43truemm \d$ \left. + \cos(2\ph)F_1(a,\r0) +
\cos(4\ph)F_2(a,\r0) + \dots \ \right\} \ , $ \medskip

\nin where, from equation (C8): \medskip

\d$ F_1(a,\r0) \= + {1\over2}{3\over2!}\arrhoq^2 +
{1\over2}{105\over4!}\arrhoq^4 + $ \par \hskip34truemm \d$ +
{15\over32}{10\b395\over6!}\arrhoq^6 + {7\over16}{2\b027\b025\over8!}\arrhoq^8 +
\dots \ . $ \medskip

As before, terms up to \d$\arrhoq^{30}$ have been analytically taken in account.
\medskip

We have now to see how we can use interference phenomena to measure the velocity
of propagation of gravitation. Precisely we want, as said in sect. IIA, to get a
zero measurement of the velocity of propagation. \medskip

To this purpose we should have two different simple radiators designed in such a
way to produce on the detector two Fourier components of gravitation forces of
the same amplitude differing only in the respective phases. \medskip

We call these two simple radiators $\alpha$ whose mass, radius and distance are
\d$M_\alpha$, $a_\alpha$ and \d$r_{0\alpha}$, and $\beta$ whose mass, radius and
distance are \d$M_\beta$, $a_\beta$ and \d$r_{0\beta}$.

First we impose that \d$
\left({a_{\alpha}r_{0\alpha}\over{r_{0\alpha}^2+a^2}}\right) \=
\left({a_{\beta}r_{0\beta}\over{r_{0\beta}^2+a^2}}\right) $ to get \d$
F_1(a_\alpha,r_{0\alpha}) \= F_1(a_\beta,r_{0\beta}),$ and eventually
\d${M_\alpha\over\ds{r_{0\alpha}^2}} \= {M_\beta\over\ds{r_{0\beta}^2}}, $
because we have to compare the forces and not merely the potentials. \medskip

In this way the values of the relevant Fourier components will essentially
depend only from the phases \d$\ph_{_{\ds\alpha}}$ and \d$\ph_{_{\ds\beta}}$.
\medskip

Suppose that the two simple radiators produce on the detector two Fourier
components of the same amplitude $A$ and phases respectively
\d$\ph_{_{\ds\alpha}}$ and \d$\ph_{_{\ds\beta}}$. Let
\d$\ph_{_{\ds\beta}}\=\ph_{_{\ds\alpha}}+{\pi\over2}+\Delta\ph$, and suppose
$m=1$ for the considered Fourier component, then the total \ut{force} acting
(with the correct frequency) on the detector will be:

$$ A \{ \cos2\ph_{_{\ds\alpha}} + \cos2\ph_{_{\ds\beta}} \} \= A \left\{
\cos2\ph_{_{\ds\alpha}} + \cos\left(2\ph_{_{\ds\alpha}}+\pi+2\Delta\ph\right)
\right\} \ , $$

\nin and it will be zero if \d$\Delta\ph=0$. \medskip

We suppose that the detector is a harmonic oscillator with \ut{negligible}
\ut{damping}, and therefore that it will resonate exactly at a very precise
frequency. This can allow the ``zero measurement" of which in section IIA and it
is on the other hand grounded on the remark I of section IIC. \medskip

As a consequence of the preceding consideration we suggest that the measurement
setup should consist of one detector and two radiators, one small ($\alpha$)
near the detector and the bigger ($\beta$) at a greater distance. \medskip

The constants $a, \r0, M$ of the two radiators should be chosen in in such a way
that the respective Fourier amplitudes of the gravitational forces be equal.
\medskip

We shall work in mode $m=1$, and, if \d$\ph_{_{\ds\alpha}}$ is the potential of
the near radiator $\alpha$ and \d$\ph_{_{\ds\beta}}$ that of the far radiator
$\beta$, we have, \ut{supposing} \ut{that} \ut{the} \ut{velocity} \ut{of}
\ut{propagation} \ut{is} $\vg$:

$$ \cases{ \ph_{_{\ds\alpha}}= 2\pi\nu{t} -{\ds{\pi}\over\ds{2}}
-{\ds{\Dphialpha}\over\ds{2}} \ , \cr
\ph_{_{\ds\beta}}=2\pi\nu\left(t-{\ds{{r}_{0\beta}}\over\ds{\vg}}\right) \ . \cr
} $$

Then: \medskip

\d$ \phantom{\cos}2\ph_{_{\ds\alpha}} \= 4\pi\nu{t} - \pi - \Dphialpha \ ,
\hskip2truecm \Dphialpha \ll 1 \ , $ \medskip

\d$ \phantom{\cos}2\ph_{_{\ds\beta}} \=
4\pi\nu\left(t-{r_{0\beta}\over{\vg}}\right) \= 4\pi\nu{t} -
{4\pi\nu{r}_{0\beta}\over{\vg}} \ , $ \medskip

where \d${4\pi\nu{r}_{0\beta}\over{\vg}}$ is the shift that should be measured
(clearly we expect \d$\vg=c$) \ ; \medskip

\d$ \cos2\ph_{_{\ds\alpha}} \= - \cos\left(4\pi\nu{t}-\Dphialpha\right)
\= - \cos4\pi\nu{t} \cos\Dphialpha - \sin4\pi\nu{t} \sin\Dphialpha \= $ \medskip
\d$ \phantom{\cos2\ph_{_{\ds\alpha}}}
\= - \cos4\pi\nu{t} \left\{1+{\cal{O}}\left(\Dphialpha\right)^2\right\} -
\sin4\pi\nu{t} \sin\Dphialpha \ , $ \medskip

\d$ \cos2\ph_{_{\ds\beta}}
\= + \cos\left(4\pi\nu{t}-{4\pi\nu{r}_{0\beta}\over\vg}\right)
\= $ \medskip \d$ \phantom{\cos2\ph_{_{\ds\alpha}}}
\= + \cos4\pi\nu{t} \cos{4\pi\nu{r}_{0\beta}\over\vg} - \sin4\pi\nu{t}
\sin{4\pi\nu{r}_{0\beta}\over\vg}
\= $ \medskip \d$ \phantom{\cos2\ph_{_{\ds\alpha}}}
\= + \cos4\pi\nu{t}
\left\{1+{\cal{O}}\left({4\pi\nu{r}_{0\beta}\over\vg}\right)^2\right\} -
\sin4\pi\nu{t} \sin{4\pi\nu{r}_{0\beta}\over\vg} \ ; $ \medskip

\d$ \cos2\ph_{_{\ds\alpha}} + \cos2\ph_{_{\ds\beta}} \= + \sin4\pi\nu{t} \left(
\sin{4\pi\nu{t}\over\vg} - \sin\Dphialpha \right)
+ {\cal{O}}\left(\Dphialpha\right)^2
+ {\cal{O}}\left({4\pi\nu{r}_{0\beta}\over\vg}\right)^2
\ . $ \medskip

It is very important to remark that in the experimental condition which we
suggested, we can make a capital use of the fact that the energy of the detector
will increase with the square of the time of irradiation (see D8) (at least
until the \ut{damping} of the \ut{detector} is negligible). This fact will allow
the detection of a very small signal. \medskip

If we measure the value of \d$\Dphialpha/2$, for which the Fourier component
$m=1$ of the motion of detector is equal zero, we have the velocity of
propagation (presumably $c$). \medskip

We can then summarize the measurement procedure in the following way. There are
two clearly distinct steps. \medskip

Let us write the Fourier component for $m=1$ of the signal of the radiator
$\alpha$ as \d$ A_\alpha\sin\ph_\alpha \ , $ and that of the radiator $\beta$ as
\d$ A_\beta\sin\ph_\beta \ . $ \medskip

The first step consists in obtaining with maximum possible precision that \d$
A_\alpha \= A_\beta \ . $ To this purpose we measure the total signal on the
detector as a function of the initial phase \d$\ph_{\alpha0}$. We modify by
successive attempts for instance \d$M_\alpha$, until the minimum of the signal
becomes equal zero. When this condition is obtained we have \d$A_\alpha \=
A_\beta \ . $ \medskip

The second step consists then in trying to measure \d$ \ph_{\alpha0} - \pi/2 $
with the maximum precision. \medskip

Should the retardation be zero (action at distance!), then \d$ \ph_{\alpha0} -
\pi/2 \= 0 \ , $ but in the reality \d$ \ph_{\alpha0} - \pi/2 $ is different
from zero, and its measure (as already pointed out in sect IIA) will give the
possibility of determining the velocity of propagation of the gravitation field.
\bigskip

\item{\bf II F} {\bf A numerical example} \medskip

We give now a numerical example for the case of a preliminary experiment (see
concluding remarks). \medskip

Let us recall (E1). We have: \medskip

\d$ (E2) \hskip1truecm {\cal{F}}_1 \= - M_{_D} {\partial\over\partial\r0} \
\left\{-2K{M\over\sqrt{\r0^2+a^2}} \ F_1(a,\r0)\right\} \= $

\d$ \= - M_{_D} {\partial\over\partial\r0} \
\left\{-2K{M\over\sqrt{\r0^2+a^2}} \ \left[
{1\over2}{3\over2!}\arrhoq^2 + {1\over2}{105\over4!}\arrhoq^4 + \right. \right.
$ \par \hskip34truemm \d$ \left. \left. + {15\over32}{10\b395\over6!}\arrhoq^6 +
{7\over16}{2\b027\b025\over8!}\arrhoq^8 + \dots \right] \right\} \= $ \medskip

\d$ \= \ - 2K{MM_{_D}\over\r0^2} \ \times \ \left\{ \
\left({\r0\over\rho}\right)^3 \ \left[ {15\over4} \arrhoq^2 + {315\over16}
\arrhoq^4 + \right. \right. $ \par \hskip53truemm \d$ \left. \left. +
{45045\over512} \arrhoq^6 + {765765\over2048} \arrhoq^8 + \dots \right] \ +
\right. $ \par \hskip32truemm \d$ \left. \ - \
\left({\r0\over\rho}\right)\phantom{^3} \ \left[ {3\over2} \arrhoq^2 +
{35\over4} \arrhoq^4 \right. \right. $ \par \hskip53truemm \d$ \left. \left. +
{10395\over256} \arrhoq^6 + {45045\over256} \arrhoq^8 + \dots \right] \right\} $
\medskip

As before, terms up to \d$\arrhoq^{30}$ have been analytically taken in account.
\medskip

Let it be: \medskip

\d$M\= 10^6 \ {\rm{gr}} \= 10^3 \ {\rm{kg}} \ , $

\d$M_{_D}\= 10^5 \ {\rm{gr}} \= 100 \ {\rm{kg}} \ , $

\d$a \ \= 100 \ {\rm{cm}} \= 1 {\rm{m}} \ , $

\d$\r0 \ \= 500 \ {\rm{cm}} \= 5 {\rm{m}} \ , $ \medskip

We have then: \medskip

$$ {\cal{F}}_1 \= - {\partial\over\partial\r0} \
\left\{-2K{MM_{_D}\over\sqrt{\r0^2+a^2}} \ F_1(a,\r0)\right\} \simeq \ -4.943
\times 10^{-3} \ {\rm{dyn}} \= -4.943 \times 10^{-8} \ {\rm{N}} \ . $$

For the considerations of the end of sect. IID it seems then not to be
impossible to perform the experiment.

\bigskip

\item{\bf III} {\bf Concluding remarks} \medskip

Supposing that the preceding considerations are not all nonsense, it is my
opinion that the next step should be experimental in nature and it should
concern the detector. \medskip

Of course one or two radiators of modest size should be prepared to test the
detectors. But they are the first objects to be designed in detail. \medskip

The singular qualities that the detectors should have require a preliminary
research that shouldn't be however too costly. Only if it will be possible to
reach a favourable conclusion, it would be reasonable to think to the more
substantial part of the investigation. \medskip

In order to carry out this task, I suggest to build a radiator \d$R_\tau$
consisting of two ``identical" simple radiators. \medskip

These radiators, that we call ``arms" \d$R_\alpha$ and \d$R_\beta$ of
\d$R_\tau$, will rotate coaxially. The angle \d$\th_{\alpha,\beta}$ between
\d$R_\alpha$ and \d$R_\beta$ shall be adjustable and measurable with precision.
\medskip

In general, if the frequency of rotation is $\nu$, the frequency of the
irradiated field will be $2\nu$, but when \d$\th_{\alpha,\beta} \= {\pi\over2}$
it will be $4\nu$. \medskip

The test of this fact shall be our task. \bigskip

\nin {\bf Acknowledgments} \bigskip

I want to thank S. Turrini for his helpful assistance in writing this work.
Discussions with S. Bergia, E. Picasso, L.A. Radicati and E. Remiddi are also
gratefully acknowledged. \medskip

Eventually I want to thank A. Passaro, Md dr, who, taking care of my health,
provided me with the mental and physical resources needed to carry out this
work.

\vskip1truecm

\nin {\bf References:} \medskip
\item{[1]} A.Einstein: Annalen der Physik, 49, 1916.
\item{[1']} A.Einstein: Les fondements de la th\'eorie de la Relativit\'e
           G\'en\'erale
\item{}    (trad. par M.Solovine), 7--71, Hermann \'Ed., Paris 1933.
\item{[2]} G.Veneziano: Il Nuovo Cimento 57A (1968), 190.
\item{[3]} B.Ferretti: Lettere al Nuovo Cimento 40, 169 (1984).
\item{[4]} B.Ferretti: Rend. Fis. Acc. Lincei \ s9, v1:281--284 (1990).
\item{[5]} B.Ferretti: Rend. Fis. Acc. Lincei \ s9, v4:193--199 (1993).
\item{[6]} See, for instance, the report in
           http://www.roma1.infn.it/rog/general/cirg/html
\item{[7]} S.Kopeikin, E.Fomalont: Proc. 6th Europ. VLBI Netw. Symp.
\item{}    June 25th-28th 2002, Bonn, Germany.
\item{[8]} S.Kopeikin: Post--Newtonian Treatment of the VLBI Experiment
\item{}    on September 8, 2002.
\item{[9]} P. Astone et al.: Zeitschrift f\"ur Physik C, 50, 21--29 (1991).
\item{[10]} P. Astone et al.: European Physical Journal C, 5, 651--664 (1998).
\item{[11]} J.A.M. Vermaseren "Symbolic Manipulation with FORM", version 2,
\item{}     CAN, Amsterdam, 1991; New features of {\tt FORM}, [math-ph/0010025].

\bye